\documentstyle[prd,aps,epsfig]{revtex}
\newcommand{\beq}{\begin{equation}}
\newcommand{\beqn}{\begin{eqnarray}}
\newcommand{\eeq}{\end{equation}}
\newcommand{\eeqn}{\end{eqnarray}}

\begin{document}
\draft
\twocolumn[\hsize\textwidth\columnwidth\hsize\csname
@twocolumnfalse\endcsname

\preprint{HUTP-97/A107; hep-ph/9801307}
\title{Larger Domains from Resonant Decay of Disoriented Chiral
Condensates}
\author{David I. Kaiser}
\address{Lyman Laboratory of Physics, Harvard University, Cambridge, MA
02138 USA}
\date{January 1998; revised November 1998}
\maketitle
\begin{abstract}
The decay of disoriented chiral condensates into soft pions is considered
within the context of a linear sigma model.  Unlike earlier analytic
studies, which focused on the production of pions as the sigma field
rolled down toward its
new equilibrium value, here we focus on the amplification of
long-wavelength pion modes due to parametric resonance as the sigma field
oscillates around the minimum of its potential.  This process can create
larger domains of pion fluctuations than the usual spinodal decomposition
process, and hence may provide a viable experimental signature for chiral
symmetry breaking in relativistic heavy ion collisions; it may also better 
explain physically the large growth of domains found in several
numerical simulations. 
\end{abstract}
\pacs{PACS 25.75.+5 \hspace*{0.2cm} Preprint HUTP-97/A107, hep-ph/9801307}
]

 
\indent  Experiments at the Relativistic Heavy Ion Collider at Brookhaven
and at the Large Hadron Collider at CERN may soon be able to probe many
questions in strong-interaction physics which have until now been studied
only on paper or simulated on a lattice.  One major area of study concerns
the QCD chiral phase transition.  In relativistic heavy ion collisions, it
is possible that non-equilibrium dynamics could produce \lq\lq disoriented
chiral condensates" (DCCs), domains in which a particular direction of the
pion field develops a non-zero expectation value.\cite{dcc1,rw,ns}
These
domains would then decay to the usual QCD vacuum by radiating soft pions.
Preliminary searches for DCCs by the MiniMax Collaboration in $p
\overline{p}$ collisions at
Fermilab have thus far not found evidence for the production and decay of
DCCs \cite{minimax}, though they are far more likely to be
created in upcoming heavy ion collisions.  Thus, understanding their
possible formation and likely
decay signatures in anticipation of further experimental work is of key
importance. \\
\indent  If these domains grow to sufficient size (on the order of 3 - 7
fm), such an experimental event would be marked by a particular clustering
pattern:  some regions within the detector would measure a large number of
charged pions but few neutral pions, while other regions of the detector
would measure predominantly neutral pions with few charged pions.\cite{rw}
Defining ${\cal R}$ to be the ratio of neutral pions to total pions,
${\cal R} \equiv n_{\pi^o}/(n_{\pi^o} + n_{\pi^+} + n_{\pi^-})$, it has
been demonstrated that the probability for measuring various ratios ${\cal
R}$ in DCC events obeys $P ({\cal R}) = (4 {\cal R})^{-1/2}$, which,
especially for small-${\cal R}$, may be easily distinguished from the
isospin-invariant result of $P({\cal R}) \rightarrow
\delta( {\cal R} - 1/3 )$. \cite{rw}  (Detecting the decay of such
DCCs could be improved by measuring the two-pion correlation functions
\cite{hirmin}, and from enhanced dilepton and photoproduction \cite{boyandcc},
in addition to studying the fraction of neutral pions produced.)
The production and subsequent relaxation
of such DCCs may also explain the so-called \lq\lq Centauro" high-energy
cosmic ray events, in which very large numbers of charged pions are
detected with only very few neutral pions. \cite{rw,cent} \\
\indent  However, as emphasized in \cite{ggp94,gm94}, if the disoriented
domains do not grow to such large scales within heavy ion collisions, such
experimental signatures
become less and less easy to distinguish from the isospin-invariant case.
Even if DCCs are produced following a heavy ion collision, if the domains
do not grow to be \lq\lq large" (that is, several fm), then the
detector would sample so many of these discrete domains within a given
run, each of which with the pion field aligned along some random
direction, that the clustering effects would be washed out.  A crucial
question, then, is whether or not sufficiently large domains might grow in
the nonequilibrium aftermath of a heavy ion collision.\\
\indent  Several authors have considered the amplification of
long-wavelength pion modes from the decay of DCCs in the context of a
linear sigma model.\cite{rw,boyandcc,ggp94,gm94}   The relevant
degrees of
freedom are modeled by the scalar fields $\sigma$ and $\vec{\pi}$, which
may be grouped together as $\Phi = (\sigma,
\vec{\pi})$.
Above the critical temperature, when the system is in the
chirally-symmetric state, the effective potential for $\Phi$ is $O (4)$
symmetric, and $\langle \Phi \rangle = 0$.  To model a
strongly-nonequilibrium situation, Rajagopal and Wilczek considered a
quench:  as the quark-gluon plasma in the interaction region between the
colliding nuclei expands and cools, the effective temperature may fall
quickly to $T \ll T_c$.  Because the zero-temperature potential is {\it
not} chirally-symmetric, domains form, and it takes some time for the
fields to evolve from $\langle \sigma \rangle = \langle \vec{\pi} \rangle
= 0$ to the new equilibrium values, $\langle \sigma \rangle \neq 0$,
$\langle \vec{\pi} \rangle = 0$.  Following the quench, the fields relax
to these new equilibrium values according to the effective potential,
$V(\Phi) = {\lambda \over 4} ( \Phi^2 - v^2)^2 - H\sigma$; that is, the
$\sigma$ field \lq rolls down' from $\sigma \sim 0$ to $\sigma \sim v$.
\\
\indent  Numerical simulations \cite{rw,ns} reveal a large amplification
of soft pion modes from the relaxation of the nonequilibrium plasma.
Previous authors have attempted to explain these
numerical results analytically in terms of spinodal decomposition:  during
the time that $\sigma$ rolls down toward $v$, pion modes with wavelengths
satisfying $k^2 \leq \lambda ( v^2 - \langle \Phi^2 \rangle )$ will grow
exponentially.  However, under the usual quench scenario, the time it
takes for
$\sigma$ to roll to $v$, and hence the maximum domain size for the DCCs,
remains too small to produce clear experimental signatures. Under this
scenario, domains typically remain pion-sized, $\sim 1.4$ fm. (See, {\it 
e.g.}, \cite{ggp94,gm94}.)  This physical mechanism alone therefore
remains incapable of explaining the large domains found in numerical
simulations. \\ 
\indent  Building on earlier work in \cite{mm95}, we consider here a
physically distinct process
which could produce larger domains of DCC, and hence might better
explain the
significant clustering observed in numerical simulations.  Rather than
amplification of pion modes while $\sigma$ rolls
down its potential hill, we focus on the parametric amplification of pion
modes as $\sigma$ oscillates around the minimum of its potential.  Because
this is a distinct process,
the growth of domains due to parametric resonance,
unlike the growth of domains due to spinodal decomposition, may reach
scales on the order of 3-5 fm. \\
\indent  This means of DCC decay is similar to cosmological post-inflation
reheating.
An early attempt was made to apply
the reheating formalism of \cite{KLS} to the decay of DCCs due to
parametric resonance in
\cite{mm95}.  However, the analytic tools for studying the
nonequilibrium,
nonperturbative dynamics of such resonant decays have improved since this
early work on reheating, and the earlier approximations, while at times
qualitatively informative, prove quantitatively unreliable.  Most
important, this earlier study \cite{mm95} approximated $\sigma$'s
oscillations as
purely periodic, in which case the equation of motion for the pionic
fluctuations reduces to the well-known Mathieu equation.  Ignoring the
nonlinear, anharmonic terms (such as $\lambda \sigma^4$) in the evolution
of $\sigma$ then yields the prediction of an infinite hierarchy of
resonance bands, with decreasing characteristic exponents.  Yet given the
nonlinear equation for $\sigma$, the equation of motion for the pions
reduces instead to a Lam\'{e} equation, which, in the cases of interest,
has only one single resonance band, with a different value for the
amplified modes' characteristic exponent.  As emphasized in
\cite{DB205}, these two differences combined can change dramatically
the predicted spectra from parametric resonance; to be useful in making
contact with experiments, these nonlinearities must be attended to, as
in the present study.  (Furthermore,
the authors of \cite{mm95} did not consider the size of domains created by
the parametric resonance, as considered here.)  Instead, we draw on the
more recent studies of
reheating in \cite{DB205,DK97,KLS2,DK98} to consider the question of DCCs
and their resonant decay. \\
\indent  Following \cite{rw}, we consider a quench scenario:  the
temperature of the plasma drops quickly from above the critical
temperature (with $\langle \Phi \rangle = 0$) to near zero.  The
effective Lagrangian density following the quench is given by
\beq
{\cal L} = - \textstyle{{1\over 2}} ( \partial_\mu \Phi )^2 -
\textstyle{{\lambda \over 4}} \left( \Phi^2 - v^2 \right)^2 +
H\sigma .
\label{lag}
\eeq
Here $H$ is an external field which breaks the chiral symmetry and picks
out the $\sigma$ direction as the true minimum.  The pion mass is
proportional to $H$.  The true vacuum is characterized by $\langle \Phi
\rangle = (f_\pi, \vec{0})$, where $f_\pi = 92.5$ MeV is the pion decay
constant.  In the limit as $H \rightarrow 0$, $f_\pi \rightarrow v$.  In
the following, we neglect $H$ in the resulting equations of motion, but
add by hand a pion mass $m_\pi = 135$ MeV; we also set $\lambda = 20.0$
and $v = 87.4$ MeV, which yield $m_\sigma = (2 \lambda f_\pi^2 +
m_\pi^2)^{1/2} = 600$ MeV. These standard values for the parameters are
chosen, as in \cite{rw,ggp94,gm94,mm95}, to fit low-energy pion dynamics.
\\
\indent  As a first approximation, we neglect effects due to the expansion
of the plasma.  Obviously the expansion of the plasma plays a crucial
role, at least for early times following the collision, in dropping the
temperature below the critical temperature.  
(Some work has been
done to incorporate analytically the effects of cosmological expansion in
the resonant decay of a massive inflaton \cite{KLS2}, which may be useful
in improving future analytic studies of DCCs and their decay).  We also
ignore noise and other medium-related effects on the resonance; as
demonstrated in the context of post-inflation reheating, such effects do
not generally destroy the parametric resonance, but rather enhance it.
\cite{noise}  \\
\indent  We study the nonequilibrium, nonperturbative dynamics by means of
a Hartree approximation, by writing $\sigma (t, {\bf x}) = \sigma_0 (t) +
\delta \sigma (t, {\bf x})$, and replacing $\vec{\pi}^3 \rightarrow
3 \langle \vec{\pi}^2 \rangle \> \vec{\pi}$ and $\vec{\pi}^2
\rightarrow \langle \vec{\pi}^2 \rangle$.  The vacuum expectation value
may be
written in terms of the field's associated (Fourier-transformed) mode
functions as $\langle \vec{\pi}^2
\rangle = \int d^3 k  \>\vert \vec{\pi}_k \vert^2/(2\pi)^3$.  Because
the $\delta \sigma$ fluctuations decouple from the pion modes in this
approximation, we will focus below on the pionic fluctuations. \\
\indent  Within a given DCC domain, the pion field will be aligned
along some particular direction, $\hat{n}_\pi$, in isospin space.  We will
therefore write $\vec{\pi} = \chi \> \hat{n}_\pi$.  In terms of the
dimensionless variables $\tau \equiv \sqrt{\lambda v^2} \> t$ and $\kappa
\equiv k/\sqrt{\lambda v^2}$, and the scaled field $\varphi (\tau) \equiv
\sigma_0 (t)/ v$, the 
coupled equations of motion take the form:
\beqn
\nonumber \varphi^{\prime\prime} + \left( \varphi^2 - 1  + \Sigma_\pi
\right) \varphi &\simeq& 0 , \\
\chi_k^{\prime\prime} +  \left( p^2 + \varphi^2  + \Sigma_\pi \right)
\chi_k &\simeq& 0,
\label{eom}
\eeqn
where primes denote $d/d\tau$, and we have defined
\beq
M \equiv m_\pi/\sqrt{\lambda v^2},\>\>p^2 \equiv
\kappa^2 + M^2 - 1,\>\> \Sigma_\pi \equiv \langle \vec{\pi}^2
\rangle/v^2 .
\label{eom2}
\eeq
Note that with the values of the parameters assumed here, $\sqrt{\lambda
v^2} = 390.9$ MeV, and $M^2 = 0.12$.  These equations of motion are
conformally equivalent to those for massless fields in an expanding,
spatially-open universe, and hence we may apply the techniques of
\cite{DK98} to study their solutions. \\
\indent  We are interested in the growth of $\vec{\pi}$ modes as
$\sigma_0$ oscillates around $v$. Having begun, following the quench, near
$\sigma_0 \sim 0$, $\sigma_0$ will roll down its potential hill toward
$v$.  The rolling field will at first overshoot the minimum at
$v$, and then begin oscillating around $v$.  The amplitude of these
oscillations will eventually be damped by the transfer of energy from this
oscillating zero mode into the $\vec{\pi}$
fluctuations.  For early times after these oscillations have begun,
however, the amplitude of $\sigma_0$ will remain nearly constant.  In this
strongly-coupled system, unlike in the weakly-coupled inflationary case,
the $\sigma$ field will execute only a few oscillations before settling in
to its minimum.  Yet, as we see below, even these few oscillations could
prove significant, since most particle production via parametric resonance
occurs in highly non-adiabatic bursts, when the velocity of the
oscillating field passes through zero. \cite{KLS}  Furthermore, because
the system has
been quenched from its initial, chirally-symmetric state, we assume that
$\Sigma_\pi$ is small at the beginning of $\sigma_0$'s oscillations.  (The
fact that spinodal decomposition alone cannot produce large DCC domains is
equivalent to $\Sigma_\pi$ remaining small while $\sigma$ rolls down its
potential hill.) Then we may solve the coupled equations for
early times after the oscillations have begun, and study the growth of the
fluctuations $\vec{\pi}_k$.  Because $\sigma_0$
begins oscillating quasi-periodically, certain pion modes will be
amplified due to parametric resonance. \\ 
\indent  The resonance will fade once the
backreaction term, $\Sigma_\pi$, grows to
be of the same order as the tree-level terms, such as $\varphi^2$.
 To study the behavior of the pionic
fluctuations, we solve the coupled equations of (\ref{eom}) for early
times after the beginning of $\sigma_0$'s oscillations, when $\Sigma_\pi$
may be neglected.  This lasts up to the time $\tau_{\rm end}$, determined
by $\Sigma_\pi (\tau_{\rm end}) = \overline{\varphi^2 (\tau)}$, where
an overline denotes time-averaging over a period of $\varphi$'s
oscillations. \\
\indent  Assuming that $\sigma_0$'s oscillations begin once
$\sigma_0$ reaches its inflection point, $\sigma_{\rm infl} = v/\sqrt{3}$,
it will roll
past the minimum and up to the point at which $V (\sigma) = V( \sigma_{\rm
infl})$,
before rolling back down through $v$.  This sets $\varphi_0 = \sqrt{5/3}$.
Because this definition of the initial amplitude is somewhat arbitrary,
we
study the resonance effects for $\varphi_0$ in the range $1 \leq \varphi_0
\leq \sqrt{2}$.
In the range $1 \leq \varphi_0 \leq \sqrt{2}$, $\varphi (\tau)$ oscillates
as \cite{DB205}
\beq
\varphi (\tau) = \varphi_0 \> {\rm dn} \left( \gamma
 \tau, \> \nu \right) ,
\label{cn}
\eeq
where ${\rm dn}(u, \nu)$ is the third Jacobian elliptic function,
$\gamma
\equiv \varphi_0 / \sqrt{2}$, and $\nu \equiv \sqrt{2 (1 -
\varphi_0^{-2})}$. Eq. (\ref{cn}) holds for $\tau \leq \tau_{\rm
end}$.  The ${\rm dn}$-function oscillates between a maximum at 1 and a
minimum at $(1 - \nu^2)^{1/2}$, with a period of $2 K (\nu)/ \gamma$,
where $K (\nu)$ is the complete elliptic integral of the first kind.
\cite{Absteg} \\
\indent  With $\varphi (\tau)$ oscillating as
in Eq. (\ref{cn}), the equation of motion for $\chi_k$ becomes the
Lam\'{e} equation of order one.  A solution for the pion modes $\chi_k
(\tau)$ may thus be written in the form \cite{DB205,DK97,DK98,Ince}:
\beq
U_k^{(\pm)} (\tau) = A (\tau) \exp \left( \pm \mu_\kappa (\nu) \gamma \tau
\right) .
\label{Uk}
\eeq
Here $A(\tau)$ is a periodic function, normalized to have unit amplitude,
and $\mu_\kappa (\nu)$ is the
characteristic exponent (also known as the Floquet index).  The form of
$\mu_\kappa$ depends on both $\kappa$ and $\nu$.  Clearly, whenever $Re
\left[ \mu_\kappa (\nu) \right] \neq 0$,
the coupled modes will be exponentially amplified.  The exact relation
between the $U_k$ modes and $\chi_k$ (and hence
$\vec{\pi}$) depends on the assumed initial conditions
following the nonequilibrium quench.  If we make the usual assumption,
that $\chi_k (\tau_0) = 1/\sqrt{2\omega_p (\tau_0)}$ and $\chi_k^\prime
(\tau_0) = -i \sqrt{\omega_p (\tau_0)/2}$, with $\omega_p^2 (\tau) \equiv
p^2 + \varphi^2 (\tau)$ \cite{DB205,juergen}, then the $\chi_k$ modes may be
written as a linear combination of $U_k^{(\pm)}$ as 
in \cite{DB205,DK97,DK98}.\\
\indent  In \cite{DK98}, these coupled equations were studied for the
range $\varphi_0 \geq \sqrt{2}$.  Proceeding in exactly the same way,
solutions may be found for the case $1 \leq \varphi_0 \leq \sqrt{2}$.  The
characteristic exponent has non-zero real parts only within a single
resonance band, given by
\beqn
\nonumber \left(\textstyle{{1\over 2}} - M^2 \right) 
- \textstyle{{1\over 2}} \sqrt{1 -
\varphi_0^2 \left( 2 - \varphi_0^2 \right)} \leq \kappa^2 \\
\leq \left( \textstyle{{1\over 2}} - M^2 \right) + \textstyle{{1\over 2}}
\sqrt{1 - \varphi_0^2
\left( 2 - \varphi_0^2 \right)} .
\label{range}
\eeqn
The resonance band includes modes with $k \leq m_\pi$ for all values of
$\varphi_0 \geq 1.23$, that is, even for amplitudes of the oscillating
field smaller than
$\varphi_0 = \sqrt{5/3}$.  As in \cite{DK98}, $\mu_\kappa (\nu)$ may be
written in terms of a
complete elliptic integral of the third kind.  The real part of $\mu_\kappa
(\nu)$ is plotted in Fig. 1.  Near the center of
the resonance band for a given value of $\varphi_0$, $Re \left[
\mu_\kappa \right]
\sim 0.1 - 0.3$.  Note that as in the numerical simulations of
\cite{rw,ns}, the strongest amplification (indicating greatest particle
production) occurs for $k \leq m_\pi$.  The maximum values of $Re \left[
\mu_\kappa \right]$ fall in the $k \rightarrow 0$ limit.
\begin{figure}
\rightline{}
\vskip 2cm
\centerline{\epsfig{file=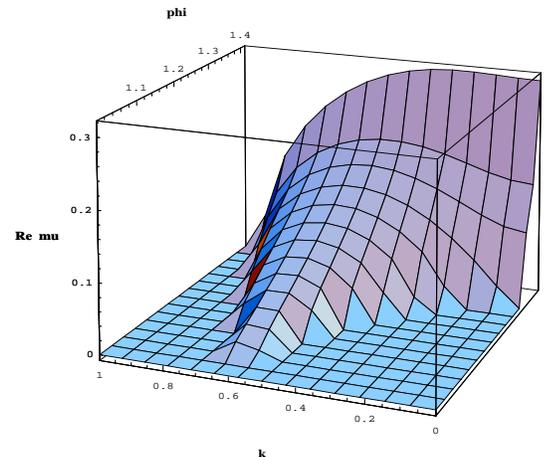,height=1.5in,width=1.8in}}
\vspace{5pt}
\caption{\small $Re \left[ \mu_\kappa (\nu) \right]$ as a function of both
the dimensionless momentum, $\kappa$, and the dimensionless initial
amplitude of $\sigma_0$'s oscillations, $\varphi_0$.  In these units,
$m_\pi = 0.35$; the largest exponents, and hence the strongest resonance,
occur for $k \leq m_\pi$.} 
\end{figure}
\indent  Given $Re \left[ \mu_\kappa \right]$, one can determine
$\tau_{\rm end}$, based on the growth of $\Sigma_\pi$.  If $\tau_{\rm
end}$ is large enough,
then observable domains of DCC could be formed and detected.  Within a
given domain, $\Sigma_\pi (\tau)$ is given as an integral over $\vert
\chi_k (\tau) \vert^2$.  To evaluate $\tau_{\rm end}$, we solve
numerically the equation
\beq
\Sigma_\pi (\tau_{\rm end}) = \int_{\rm res. \> band} \frac{d\kappa \>
\kappa^2}{2
\pi^2} \vert \chi_k (\tau_{\rm end}) \vert^2 = \overline{\varphi^2} ,
\label{tend}
\eeq
where the integral extends over the single resonance band.  The 
time-average of the (dimensionless) 
oscillating $\sigma$ field, $\overline{\varphi^2}$, may be written
in terms of $E (\nu)$,  
the complete elliptic integral of the second kind, using the integral of
${\rm dn}^2 (u,\nu)$ over
a period of its oscillations (see \cite{Absteg}).  Eq. (\ref{tend}) then
 yields $t_{\rm end} \sim 5$ fm/$c$
over most of the range $1 \leq \varphi_0 \leq \sqrt{2}$.  This should be
compared with the usual spinodal decomposition scenario, in which the
pionic fluctuations would be exponentially amplified only for the brief
period $t_{\rm spinodal} \simeq \sqrt{2}/m_\sigma \simeq 0.47$ fm/$c$. \\
\indent  In order to find the characteristic sizes to which DCC
domains may grow before the parametric resonance is damped, we may follow
\cite{dbdomain} and evaluate the two-point correlation
function:
\beqn
\nonumber D (t,\> {\bf r}) &\equiv& \langle \vec{\pi} (t, \>{\bf r})
\vec{\pi} (t,\>
{\bf 0}) \rangle = \int
\frac{d^3 k}{(2\pi)^3} e^{i {\bf k} \cdot {\bf r}} \vert \vec{\pi}_k
\vert^2 \\
&=& (\lambda v^2)^{-3/2} \int\frac{d \kappa \> \kappa^2}{2\pi^2} j_0
(\kappa x) \vert \chi_k (\tau) \vert^2 ,
\label{twopoint}
\eeqn
with the dimensionless length defined by ${\bf x} \equiv \sqrt{\lambda
v^2} \> {\bf r}$.  We have also used $\vec{\pi}_k (t) = \chi_k (\tau)
\> \hat{n}_\pi$ within a given domain.  We may solve this integral in the
saddle-point approximation, making use of Eq. (11.4.29) of
\cite{Absteg}, with the result that
\beqn
\nonumber D (t, \> {\bf r}) &\propto& \Sigma_\pi (t) \exp \left( - r^2/
\xi_D^2 (t) \right) , \\
\xi_D^2 (t) &\equiv& 4 \gamma \vert Re\left( \partial^2 \mu_\kappa (\nu) /
\partial \kappa^2 \right) \vert_{\kappa_{\rm max}} (t
/ \sqrt{\lambda v^2} ) .
\label{xi}
\eeqn
Eq. (\ref{xi}) reveals that the domain size grows as $t^{1/2}$.  The
maximum correlation length, $\xi_D (t_{\rm end})$, is plotted in Fig. 2.
\begin{figure}
\vspace{20pt}
\centerline{\epsfig{file=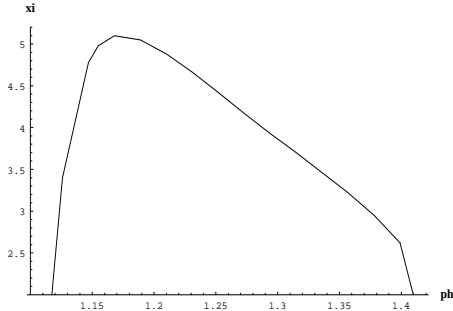,height=1in,width=1.5in}}
\vspace{7pt}
\caption{\small Maximum domain size for DCCs, $\xi_D (t_{\rm end})$, as a
function of $\varphi_0$, in units of fm.}
\end{figure}
Over much of the range $1 \leq \varphi_0 \leq \sqrt{2}$, $\xi_D (t_{\rm
end})$ lies between 3 - 5 fm.  The usual spinodal
decomposition process, in the absence of the \lq annealing' studied in
\cite{gm94}, on the other hand, can only create domains of order
1.4 fm. \cite{ggp94}.  If \lq annealing' is effective, domains from
spinodal decomposition may grow to $3 - 4$ fm; yet independent of the
dynamics as $\sigma$ rolls down its potential hill, it still must end
its evolution by oscillating around $v$, and the resonant production of
pions would follow as studied here. \\
\indent  Parametric resonance offers a promising means of producing
observable signals from the production and decay of disoriented chiral
condensates in the aftermath of relativistic heavy ion collisions.  Unlike
spinodal decomposition alone, this physical process may explain the
signficant growth of DCCs found in numerical simulations.  Future analytic
studies should better include effects from the
expansion of the plasma, and from $\delta \sigma - \vec{\pi}$ and
$\vec{\pi} - \vec{\pi}$ scatterings, which are neglected here in the
Hartree approximation.  If these effects remain subdominant, however, then
the resonant decay of DCCs should produce low-momentum pions with a
distribution observably distinct from the isospin-invariant case.\\
%
\indent  It is a pleasure to thank Krishna Rajagopal, Dan Boyanovsky,
Stanislaw Mrowczynski, and Hisakazu Minakata for helpful discussions.
This research was supported in part by NSF grant PHY-98-02709.

\end{document}